\documentstyle [12pt]{article}
\begin{document}

\begin{titlepage}
\rightline{ULB-PMIF/93-02}
\vspace{1cm}
\begin{centering}

{\Large Ghosts of ghosts for second class constraints}

\vspace{2cm}

{\large  Marc Henneaux}\\
\vspace{1cm}
Facult\'e des Sciences, Universit\'e Libre de Bruxelles,\\
Campus Plaine C.P. 231, B-1050 Bruxelles, Belgium\\
and\\
 Centro de Estudios Cient\'{\i}ficos de Santiago,\\
 Casilla 16443, Santiago 9, Chile\\

\vspace{2.5cm}
{\large Abstract}
\vspace{.5cm}

\end{centering}

When one uses the Dirac bracket, second class
constraints become first class.
Hence, they are amenable to the BRST
treatment characteristic of
 ordinary first class
constraints. This observation is the
starting point of a recent
investigation by Batalin and Tyutin, in
which all the constraints
are put on the same footing.  However,
because second class
constraints identically vanish as operators
in the quantum theory, they are
quantum-mechanically reducible
 and require therefore ghosts of ghosts.
Otherwise, the BRST cohomology would not
yield the correct physical spectrum.
We discuss how to incorporate this feature
in the formalism
and show that it leads to an
infinite tower of ghosts of ghosts.
An alternative treatment, in which the
brackets of the ghosts are modified,
is also mentioned.

\end{titlepage}

\thispagestyle{empty}
\vfill
\pagebreak

Constraints in constrained Hamiltonian systems can
arise for different physical
reasons \cite{Dirac}.  They may originate
from a gauge invariance of the
theory, in which case they are first class.
Or they may indicate that the
$q-p$ commutation rules must be modified,
in which case they are second class.
First class constraints kill twice as many
degrees of freedom as second class
constraints do.  Geometrically, first
class constraints define
co-isotropic submanifolds, while
second class constraints define regular
submanifolds (i.e. submanifolds
with an invertible
induced two-form - for a recent
review, see \cite{Henn}).

The standard quantization rules implement
first class and second class
constraints quite differently.  This is
natural in view of the different
physical meanings of the constraints.
However, this feature is
somewhat unfortunate because it requires
an explicit separation of the constraints
into first and second classes before
going to the quantum theory.
  This is
always possible in principle, but in
practice, the split may be cumbersome
or may spoil manifest
Lorentz invariance.  For this reason,
various efforts have been devoted to the
question of quantizing the constraints
in a more uniform manner.

Recently, Batalin and Tyutin have constructed
a very interesting algebraic scheme
in which the constraints are formally
kept on the same footing \cite{Bat1,Bat2,Bat3}.  Their
starting point is the observation that all the
constraints are first class if one
uses the Dirac bracket. Hence, one may define
a BRST generator in
which one associates a conjugate pair of
ghosts to all (first {\em and} second
class) constraints. In that approach,
the ghost spectrum appears to make
no distinction between the constraints.
Batalin and Tyutin
 then go on to develop a powerful formalism that
 incorporates
the covariance properties of the Dirac
bracket under redefinitions
of the constraints, but this analysis
will not be needed for our discussion,
which focuses on
a much more elementary point. {\em We shall
only investigate here the extent to which one
can introduce a ghost spectrum uniformly
for all the constraints}.  The main
result of this letter is to indicate that
the ghost spectrum must be augmented
by ghosts of ghosts for the second class
constraints.
These ghosts of ghosts destroy the symmetry
between first class and second class
constraints. A uniform treatment may be
recovered if
one introduces further an infinite tower
of additional ghosts of ghosts.

Let $z^A$ ($A = 1, ...,n$) be the phase
space coordinates and let
$\phi_a(z^A)$ ($a = 1,...,m$) be the
constraint functions. We assume the constraints
to be irreducible for simplicity.
We denote the Dirac
bracket by $[ , ]$ and will make no
mention of the original Poisson bracket
thereafter.  One
important feature of the Dirac bracket
is that it is degenerate in the
algebra $C^{\infty}(P)$ of smooth phase
space functions $f(z^A)$.  Indeed,
if $\chi_{\alpha}$ are the second class
constraints that have been used in
the definition of the Dirac bracket, then one has
\begin{equation}
[f,\chi_{\alpha}] = 0.
\end{equation}
It is the degeneracy of the Dirac bracket that enables one to
distinguish between first class
and second class constraints when one has
eliminated the Poisson bracket.  Although
degenerate in the algebra $C^{\infty}(P)$
of smooth functions defined in
the entire phase space, the Dirac bracket
is regular in the algebra
$C^{\infty}(\Sigma_2)$ of functions defined
on the surface of the second
class constraints.  But since we want to
maintain a symmetric treatment of the
constraints, we do not reduce the system
to $\Sigma_2$ at this stage and work instead
with the full algebra $C^{\infty}(P)$.

In terms of the Dirac bracket, all
the constraints are first class
\begin{equation}
[\phi_a, \phi_b] = C^c_{ab} \phi_c
\end{equation}
and we shall denote by
$\Omega(z^A, \eta^a, {\cal{P}}_{a})$ the corresponding
BRST generator{\footnote{It
is denoted by $\Omega_0$ in \cite{Bat1}.  Note that
once the solution
of (3) is found, the general solution of the
equations of \cite{Bat1,Bat2} reads,
in the notations of \cite{Bat1,Bat2},
$$\Omega_0 + \Gamma^*_A(\Gamma^A + O(\eta^a)) + O(\Gamma^*_AJ\Gamma^*_B)$$
(and $\Delta$ is defined by Eq. (3.5) of
\cite{Bat2}).  Thus, one may say that the
heart of the construction of \cite{Bat1,Bat2} is
contained in the equation
$[\Omega_0, \Omega_0] = 0$.}}.  There is one pair
of conjugate ghosts ($\eta^a, {\cal{P}}_{a}$)
for each constraint $\phi_a$.  The
BRST generator fulfills
\begin{equation}
[\Omega, \Omega] = 0
\end{equation}
(in the Dirac bracket) everywhere in phase space and not just
on the surface of the second class constraints.
The fact that the Dirac bracket
is degenerate does not prevent one from
obtaining the BRST generator in the
usual manner.  The Koszul-Tate differential
used in the standard construction
explained in \cite{Henn} is acyclic at positive resolution
degree independently of what the bracket is,
and the first class property (2) guarantees
that one can complete the
$\phi_a \eta^a$-term in
\begin{equation}
\Omega = \phi_a \eta^a + ...
\end{equation}
in the usual recursive procedure, in such a way that (3) holds.

However, the degeneracy of the bracket has
one dramatic consequence on the cohomology.
As it has been shown in
\cite{Kimura}, the classical BRST cohomology
is not given by the longitudinal
cohomology along the gauge orbits, as in the
standard case.  It is rather given by
the longitudinal cohomology tensored by the
ghosts $\eta^{\alpha}$ of
the second class constraints, which are
associated with no gauge direction but
which are not removed in cohomology.
In the regular case with
invertible Poisson bracket, first class
constraints play a dual role.  Not only
do they constrain the system, but also,
they generate the gauge transformations.
This dual role is lost when the
bracket is degenerate, since the (original)
second class constraints appear as
first class but generate no gauge transformation.
They actually generate nothing
at all because of (1).  The
disappearance of the dual role played
by the constraints makes previous
analyses of the BRST cohomology
inapplicable.

To make this point clear, let us
consider the simple case of
a two-dimensional phase space with
coordinates $q,p$.  We take as
constraints
\begin{equation}
q = 0, \; \; p = 0.
\end{equation}
This system has no physical degree of
freedom and the observables are
just the constants since the
constraint surface is reduced to a single
point.  The Dirac bracket is
\begin{equation}
[q,p] = 0.
\end{equation}
The BRST charge reads
\begin{equation}
\Omega = q \eta^1 + p \eta^2.
\end{equation}
The gauge orbits are $0$-dimensional ($[f, \phi_a] = 0$)
so that there is no
longitudinal
one-form.  The exterior longitudinal algebra on the
constraint surface reduces to the
constants.  Yet, the BRST cohomology is
given by the polynomials in the ghosts
$\eta^a$, which cannot be interpreted as
longitudinal forms. Indeed, the ghost
momenta ${\cal P}_1$ and ${\cal P}_1$ kill
$q$ and $p$ in cohomology, but
there is nothing to remove the
ghosts.  The classical BRST cohomology is
thus too large {\footnote {It is true
that the BRST cohomology at ghost number
zero is given by the constants. One
may argue that this is physically just what is needed. But
this feature disappears in the quantum
theory, where one finds non trivial operators
at ghost number zero (see below).}}.

Let us now turn to the quantum theory.
The space of quantum states must
provide an irreducible representation
of the Dirac brackets.  Accordingly,
 each second class constraint must be
realized as the zero operator, as the
following argument shows. Each $\hat{\chi_{\alpha}}$
commutes with everything, so that
it is a multiple of the identity. Furthermore,
it must have vanishing expectation
value for physical states and hence, this
multiple of the identity must be zero.
 As stressed in \cite{Bat1}, this
is inevitable.  Thus, there
exist non-vanishing phase space
functions that are represented by the
operator $\hat{0}$, namely $\chi_{\alpha}J(z^A)$.
This feature enlarges
even further the BRST cohomology, by
introducing also the ghost momenta ${\cal P}_{\alpha}$
in the cohomology. These variables
are no longer eliminated by the constraints $\chi_\alpha$,
which become quantum-mechanically empty.

In the case of the model (5), the states
can be represented as functions of the ghosts,
$\psi = \psi(\eta^1, \eta^2)$.  The operators are
\begin{equation}
\hat{q} \psi = 0 = \hat{p} \psi, \\
\hat{\eta^a} \psi = \eta^a \psi, \\
\hat{{\cal P}_a} \psi = - i {\frac {\partial}{\partial \eta^a}} \psi.
\end{equation}
The quantum BRST generator identically
vanishes and so any operator
\\$\hat{A}(\hat{\eta}, \hat{{\cal P}})$ defines a BRST
cohomological class.  For instance,
$\hat{\eta^1} \hat{{\cal P}_2}$ is a ghost
number zero BRST invariant operator, but it does not
correspond to any of the true physical observables,
since these are the multiples
of the identity (no physical degree of
freedom for the model at hand).

Now, while there is no way to recover
the correct cohomology at
the classical level by standard means,
this is not so at the quantum level.
This is because when one passes to quantum mechanics,
all the first class constraints recover their
characteristic dual role of both
constraining the system and
generating the gauge transformations,
or {\em of doing neither at all} (but
not of doing one and not the other).
The operators $\hat{\chi_{\alpha}}$
corresponding to the
original second class constraints
do not generate any transformation, as
in the classical theory, but they also do not
constrain the system anymore since the
equations $\hat{\chi_{\alpha}} = 0$ are of
the form $ 0 = 0$.  Hence, one is in the
standard setting of the usual Hamiltonian
BRST construction, but, of course, in the
reducible situation since the constraints
$\hat{\chi_{\alpha}} = 0$ are
empty : the second class constraints
 cease to be irreducible quantum-mechanically.  .

As is well known, in order to treat this case,
one needs to introduce
one pair of conjugate
ghosts of ghosts for
each second class constraint, even though
these constraints are irreducible
classically.  The correct
quantum BRST charge is given by
\begin{equation}
\hat{\Omega} = \hat{\phi_a} \hat{\eta^a} +
\hat{{\cal P}_{\alpha}} \hat{\gamma^{\alpha}}
+ ...
\end{equation}
where ($\gamma^\alpha, \pi_\alpha$) are
the ghost of ghost conjugate
pairs. Once the spectrum includes the ghosts of ghosts,
the BRST generator (9) possesses the
correct cohomology.  Indeed, the
main motivation for introducing
ghosts of ghosts is precisely based
on cohomological considerations
\cite{Fisch}.  The case of empty constraints
is explicitly treated in \cite{Henn} (chapter 10).

In the case of the above simple model, one has
\begin{equation}
\hat{\Omega} = \hat{{\cal P}_1} \hat{\gamma^1} +
\hat{{\cal P}_2} \hat{\gamma^2}
\end{equation}
The ghost of ghost pairs kill the ghost pairs
in cohomology and the quantum
operator BRST
cohomology is correctly given by the multiples
of the identity. More precisely,
the BRST transformation reads
\begin{equation}
s \eta^1 = \gamma^1, \; s \gamma^1 = 0,
\end{equation}
\begin{equation}
s \eta^2 = \gamma^2, \; s \gamma^2 = 0,
\end{equation}
\begin{equation}
s \pi_1 = {\cal P}_1, \; s {\cal P}_1 = 0,
\end{equation}
\begin{equation}
s \pi_2 = {\cal P}_2, \; s {\cal P}_2 = 0.
\end{equation}
Hence, the pairs ($\eta^1,\gamma^1$),
($\eta^2,\gamma^2$), ($\pi_1,{\cal P}_1$),
($\pi_2,{\cal P}_2$) mutually cancel each other in cohomology.

We thus see that the approach in which
one treats all the constraints as
first class  and uses the Dirac bracket
must be supplemented by ghosts of ghosts
for the original second class constraints,
even if these are irreducible.
The ghosts of ghosts
can be incorporated by means of the standard
techniques.  However, this spoils the
uniform treatment of the constraints
since there are ghosts of ghosts only
for the second class constraints and not for
the original irreducible first class constraints.

One may try to avoid this problem by modifying the ghosts
commutation relations.  In this
tentative approach, one would replace the ghost
Poisson brackets
by ``Dirac brackets" such that the
ghost variables associated with the second
class constraints (anti)commute with everything.  For instance,
for the model (4), one would set
\begin{equation}
[\hat{\eta^a}, \hat{{\cal P}_b}] = 0,
\end{equation}
and one would represent each of the
ghost variables as the zero operator in
the quantum theory.  The space
of quantum states would then be
one-dimensional and the BRST cohomology would be
correct.  There would be no need for ghosts of ghosts.
However, since this approach requires a separation of
the ghosts respectively associated
with the first and second class constraints, it
fails to provide a uniform treatment
of the constraints in the general case.

Alternatively, one may try to keep a  uniform
treatment of the constraints at the
level of the ghosts of ghosts by
introducing ghosts of ghosts also for the
first class constraints. This is unnecessary
since these are irreducible but it can
be done if one brings in
trivial reducibility conditions among
the first class constraints, of the
form $0$ times (constraints) $=0$.
These trivial reducibility relations are empty
and thus,
needs in turn ghosts of ghosts of ghosts
at the next level \cite{Henn,Fisch}.
Since there are empty relations only for
the original first class
constraints, the procedure spoils again
uniformity.  One can cure this defect by
associating ghosts of ghosts of ghosts
also with the original second class
constraints.  This can be done
by introducing again trivial reducibility
relations at that level.  But this forces
one to go on at the next level, and so on,
so that a symmetric treatment of all the constraints
leads to an infinite tower of ghosts of
ghosts.  Because of the well known
difficulties
associated with infinite towers of ghosts
of ghosts, the prospects of reaching
a manageable uniform treatment of all the
constraints are rather grim.
This is perhaps not too surprising in view
of the very different physical
roles played by first class and second class constraints.

\vspace{2cm}
\noindent
{\bf Acknowledgements}\\

This work has been supported in part by
a grant from the F.N.R.S. and by a
research contract with
the Commission of the European Communities.
\vfill
\eject

\end{document}